\documentclass[a4paper,11pt]{article}

\makeatletter
\@addtoreset{equation}{section}

\makeatother

\usepackage{multirow}
\usepackage{booktabs}
\usepackage[dvips]{graphicx}
\usepackage{amsmath}
\usepackage{amsthm}
\usepackage{amssymb}
\usepackage{eepic}
\usepackage{amscd}
\usepackage{longtable}
\usepackage{array}

\newtheorem{theorem}{Theorem}[section]

\newtheorem{proposition}[theorem]{Proposition}

\theoremstyle{definition}

\newtheorem{remark}[theorem]{Remark}

\newcommand{\q}[1]{\left[ #1 \right]}
\newcommand{\mapsup}{  \text{\rotatebox[origin=c]{90}{$\mapsto$} }   }
\newcommand\dotsd{\text{\rotatebox[x=1mm,y=2.5mm]{270}{$\ddots$}}}
\newcommand{\mapsdown}{\text{\rotatebox[origin=c]{270}{$\mapsto$} }   }
\newcommand{\C}{{\mathbb C}}
\newcommand{\Z}{{\mathbb Z}}
\newcommand{\brabra}{{\langle\!\langle}}
\newcommand{\ketket}{{\rangle\!\rangle}}

\newcommand{\nc}{\newcommand}
\nc{\rnc}{\renewcommand}
\nc{\nn}{\nonumber}

\newcommand{\ket}{{\rangle}}
\DeclareMathOperator{\Tr}{Tr}

\begin{document}

\title{Spin-spin correlation functions of the $q$-VBS state \\
 of an integer spin model}

\author{
 Chikashi Arita\thanks{Faculty of Mathematics, Kyushu University }
\ 
and Kohei Motegi\thanks{ Okayama Institute for Quantum Physics}  }

\date{\ }

\maketitle

\begin{abstract}
We consider the valence-bond-solid ground state
 of the $q$-deformed higher-spin AKLT model ($q$-VBS state)
 with $q$ real.
We investigate the eigenvalues and eigenvectors of a matrix
($G$ matrix), which is
constructed from the matrix product representation
of the $q$-VBS state.
We compute the longitudinal
and transverse spin-spin correlation functions, and
determine the correlation amplitudes and correlation lengths.
\end{abstract}



\section{Introduction}
In one-dimensional quantum systems, a completely different behavior 
for the integer spin
chains from the half-integer spin chains was predicted by Haldane
\cite{Haldane1,Haldane2}.
The antiferromagnetic isotropic
spin-1 model introduced by Affleck, Kennedy, Lieb and Tasaki
(AKLT model) \cite{AKLT},
whose ground state can be exactly calculated, has been a useful toy model
to validate Haldane's prediction of the massive behavior
for integer spin chains.
Moreover, it lead to a deeper understanding for integer spin chains such as the
discovery of the special type of long-range order \cite{DR,Tasaki}. 
The AKLT model has been generalized to
higher-spin models,
 anisotropic models, etc
 \cite{AAH,KSZ1,KSZ2,KSZ3,TS1,BY,GR,SR,TZX,TZXLN,KM,AHQZ}.
The Hamiltonians are essentially linear combinations
of projection operators with nonnegative coefficients,
and their ground states are called valence-bond-solid (VBS) state.
Recently, the VBS state was investigated
in perspective of
its relation to quantum information and experimental implementation
by means of optical lattices,
see Refs.~\cite{VMC,GMC}  for example.

There are largely three types of representations for the ground state which
are equivalent to each other:
the Schwinger boson representation, the spin coherent
representation and the matrix product representation.
For isotropic higher-spin models,
the spin-spin correlation functions \cite{FH}
and the entanglement entropy \cite{XKHK,KX}
have been calculated by utilizing the spin coherent representation
and the properties of Legendre polynomials.
For the $q$-deformed spin-1 model,
spin-spin correlation functions were evaluated 
\cite{KSZ1,KSZ2,KSZ3}
 from the matrix product representation.
 
In this paper, we consider the ground state of
 a $q$-deformed higher-integer-spin model
 which was constructed recently in Ref.~\cite{M} ($q$-VBS state).
From its matrix product representation,
we analyze one and two point functions of the $q$-VBS
ground state.
We notice that a matrix, which is constructed from the matrix product
representation, plays a fundamental role in computing correlation functions,
especially spin-spin two point correlation functions.
Investigating the structure of the matrix in detail,
we obtain its eigenvalues and eigenvectors.
Utilizing the results, we determine the correlation
amplitudes and correlation lengths of the longitudinal and transverse
spin-spin correlation functions.

This paper is organized as follows.
In the next section, we briefly review
the quantum group $U_q(su(2))$, and investigate the
finite dimensional highest weight representation in terms of
Schwinger bosons.
In Section \ref{Sec:qVBS},
  we precisely define the higher-spin generalization of the
  $q$-deformed AKLT model on an $L$-site chain, and
  rigorously derive its $q$-VBS ground state
  in a matrix product form.
The squared norm of the state will be written
  in terms of the trace of the $L$-th power of a matrix $G$,
  which plays an important role in this paper.
In section \ref{Sec:spec-of-G},
 we obtain the eigenvalues and eigenvectors of $G$.
Utilizing them,
  we compute one and two point functions
  in Section \ref{Sec:correlation}.
Especially, we determine the correlation amplitudes
and correlation lengths of the longitudinal and transverse
spin-spin correlation functions. 
Section \ref{Sec:Conclusion}
   is devoted to the conclusion of this paper.

\section{\label{Sec:} The quantum group $U_q(su(2))$ }
We introduce several notations,
  fixing a real number $q$ throughout this paper.
Let us define the $q$-integer, $q$-factorial
 and $q$-binomial coefficient for $N\in\Z_{\ge 0}$ as
\begin{align}
\begin{split}
    [N]=\frac{q^N-q^{-N}}{q-q^{-1}}, \quad
    [N]!=
      \begin{cases}
        \displaystyle   \prod_{I=1}^{N} [I]  & N\in {\mathbb N},\\
           1   &   N=0,
      \end{cases}     
 \\
   \left[ \begin{array}{c}  N \\ K \end{array} \right]
       =\begin{cases}
        \displaystyle   \frac{ [N]! }{ [K]! [N-K]! } & K=0,\dots,N,\\
           0   &   \rm otherwise,
      \end{cases}       
\end{split}
\end{align}
respectively.

The quantum group $U_q (su(2))$ \cite{Drinfeld,Jimbo}
  is defined by generators $X^+,X^-$ and $H$
  with relations
\begin{align}
  \left[X^+,X^-\right] = \frac{q^H-q^{-H}}{q-q^{-1}},\quad
  \left[H,X^{\pm}\right] =\pm 2 X^{\pm} .
\end{align}
The comultiplication is given by
\begin{align}
  \Delta \left( X^\pm \right)
      =  X^\pm \otimes q^{H/2} +  q^{-H/2} \otimes X^\pm,    \quad
  \Delta (H)  = H\otimes \mathrm{Id} + \mathrm{Id} \otimes H  .
\end{align}

$U_q (su(2))$ has the Schwinger boson representation,
where the generators are realized as
\begin{align}
  X^+=a^\dagger b,\quad X^-=b^\dagger a ,\quad H=N_a-N_b,
\end{align}
with $q$-bosons $a$ and $b$ satisfying
\begin{align}
   &a a^\dagger -q a^\dagger a = q^{-N_a},  \quad
   b b^\dagger -q b^\dagger b = q^{-N_b},  \\
   &[N_a,a] = -a ,\quad [N_a,a^\dagger] = a^\dagger,  \quad
   [N_b,b] = -b ,\quad [N_b,b^\dagger] = b^\dagger.
\end{align}
We denote the space where $(2j+1)$-dimensional
highest weight representation of $U_q(su(2))$ is realized by $V_j$.
The basis of $V_j$ is given by
\begin{align}
|j;m \ket=\frac{(a^{\dagger})^{j+m}(b^{\dagger})^{j-m}}
{ \sqrt{ [j+m]! [j-m]! }}
|\mathrm{vac} \ket, \ \ (m=-j, \dots, j). \label{basis}
\end{align}

The Weyl representation which we describe below,
is an equivalent representation to
the Schwinger boson representation, 
and is efficient for practical calculation.
Let us denote the $q$-bosons $a$ and $b$ acting on the 
$\alpha$-th site as $a_\alpha$ and $b_\alpha$.
The Weyl representation is to represent
$a_\alpha^\dagger$, $b_\alpha^\dagger$,
$a_\alpha$ and $b_\alpha$ on the space of polynomials
$\mathbb{C}[x_\alpha,y_\alpha]$ as
\begin{align}
a_\alpha^\dagger = x_\alpha,\quad
b_\alpha^\dagger = y_\alpha,\quad 
a_\alpha = \frac{1}{x_\alpha}\frac{D_q^{x_\alpha}-D_{q^{-1}}^{x_\alpha}}{q-q^{-1}},
\ \
b_\alpha=\frac{1}{y_\alpha}\frac{D_q^{y_\alpha}-D_{q^{-1}}^{y_\alpha}}{q-q^{-1}},
\end{align}
where
\begin{align}
D_{p}^{x_\alpha}f(x_\alpha,y_\alpha)=f(px_\alpha,y_\alpha),
\ \ D_{p}^{y_\alpha}f(x_\alpha,y_\alpha)=f(x_\alpha,py_\alpha).
\end{align}
The generators of $U_q(su(2))$ are now represented as
\begin{align}
   X^+_\alpha = \frac{x_\alpha}{y_\alpha}
   \frac{ D_q^{y_\alpha}  - D^{y_\alpha}_{q^{-1}} }
     { q-q^{-1} },\ 
   X^-_\alpha = \frac{y_\alpha}{x_\alpha}
   \frac{ D_q^{x_\alpha}  - D^{x_\alpha}_{q^{-1}} }
     { q-q^{-1} },\ 
   q^{H_\alpha} = D^{x_\alpha}_q D^{y_\alpha}_{q^{-1}}.
\end{align}

The tensor product of two irreducible representations has the
Clebsch-Gordan decomposition
\begin{align}
  V_{S} \otimes V_{S} &= \bigoplus_{J=0}^{2S}V_J , \\
\label{CGdeco}
  |S;m_1\rangle \otimes|S;m_2\rangle
&= \sum_{J=0}^{2S}
 \q{\begin{array}{ccc}
    S   & S    & J    \\
    m_1 & m_2 & m_1+m_2
   \end{array}}
|J; m_1+m_2\rangle  ,
\end{align}
where
\begin{align}
&\q{\begin{array}{ccc} S_1 & S_2 & J \\ m_1 & m_2 & m \end{array}}
= \delta_{m_1+m_2,m}(-1)^{S_1-m_1}
q^{m_1(m_1+m_2+1)+ \{ S_2(S_2+1)-S_1(S_1+1)-J(J+1) \} /2} \nn \\
&\times \sqrt{ \frac{[J+m]! [J-m]! [S_1-m_1]! [S_2-m_2]! [S_1+S_2-J]! [2J+1]
}{[S_1+m_1]! [S_2+m_2]! [S_1-S_2+J]! [S_2-S_1+J]! [S_1+S_2+J+1]! } }
\\
&\times \sum_{z=\mathrm{Max}(0,-S_1-m_1,J-S_2-m_1)}^
{\mathrm{Min}(J-m,S_1-m_1,S_2+J-m_1)}
\frac{ (-  q^{m+J+1} )^z [S_1+m_1+z]! [S_2+J-m_1-z]!}{[z]! [J-m-z]! [S_1-m_1-z]! [S_2-J+m_1+z]!}, \nn
\end{align}
is the $q$-analog
 of the Clebsch-Gordan coefficient \cite{KR}.
(The factor $q^{m_1 m_2/2}$ is missing in Ref.~\cite{KR}.)
This coefficient 
  is compatible with the inverse
  of the decomposition \eqref{CGdeco}
\begin{align}
  |J;m\rangle = \sum_{m_1+m_2=m}
  \q{\begin{array}{ccc}
    S   & S    & J    \\
    m_1 & m_2 & m_1+m_2
   \end{array}}
 |S;m_1\rangle \otimes|S;m_2\rangle .
\end{align}

For later purpose, we will also
  investigate the Clebsch-Gordan decomposition
  of $U_q(su(2))$ in terms of the Schwinger boson
  or the Weyl representation.
Utilizing
\begin{align}
  \Delta X^{\pm}_{\alpha\beta}
 =& X^\pm_\alpha \otimes q^{H_\beta /2}
      + q^{-H_\alpha /2} \otimes X^\pm_\beta,
\end{align}
one can show that the highest weight vector
  $v_J \in V_J \ (   \Delta X^+ v_J  = 0)$
acting on the $\alpha$-th and $\beta$-th sites
 is given by
\begin{align}
  v_J = (x_\alpha x_\beta)^J
     \prod_{\nu=1}^{2S-J}
      (x_\alpha y_\beta -q^{2(\nu-S-1)}x_\beta y_\alpha ).
\end{align}
Moreover, we can show the following:
\begin{proposition}\label{X^-^nv_J}
\begin{align}
\begin{split}
  \left( \Delta X^-_{\alpha\beta} \right)^n v_J
  =& (x_\alpha x_\beta)^{J-n} q^{nS} [n]!
     \sum_{\mu=0}^n  q^{-2\mu S}
       \q{\begin{array}{c} J \\ \mu \end{array} }
       \q{\begin{array}{c} J \\ n-\mu \end{array} }
     \left(x_\alpha y_\beta \right)^\mu
     \left(x_\beta y_\alpha \right)^{n-\mu} \\
&    \times
     \prod_{\nu=1}^{2S-J}
      \left( x_\alpha y_\beta -q^{2(\nu-S-1)}x_\beta y_\alpha \right) .
\end{split}
\label{vectorexpression}
\end{align}
\end{proposition}

A proof of this proposition 
  is given in Appendix \ref{Sec:proof}.

\begin{remark}
Let $n\ge 2J+1$.
Noting
\begin{align}
  \q{\begin{array}{c} J \\ \mu \end{array}}
  \q{\begin{array}{c} J \\ n-\mu \end{array}}
  =0,
\end{align}
for $0 \le \mu\le n$,
one can see that
\begin{align}
  \left( \Delta X^-_{\alpha\beta} \right)^n v_J = 0.
\end{align}
\end{remark}

\section{\label{Sec:qVBS}$q$-VBS state}
The model we treat in this paper
 is an anisotropic integer spin-$S$ Hamiltonian
 on an $L$-site chain with the periodic boundary condition
\begin{align}
&{\mathcal H}=\sum_{k\in \Z_L} 
 \sum_{J=S+1}^{2S}C_J(k,k+1) \left(\pi_J\right)_{k,k+1},
\label{hamiltonian}
\end{align}
where $C_J(k,k+1) > 0$, and $\left(\pi_J\right)_{k,k+1}$,
which acts on the $k$-th and $(k+1)$-th sites,
is the $U_q (su(2))$ projection operator
from $V_S \otimes V_S$ to $V_J$ as
\begin{align}
\begin{split}
\pi_J=&
\sum_{m_1,m_2, m_1', m_2'=-S}^S
\q{\begin{array}{ccc} S & S & J \\ m_1 & m_2 & m_1+ m_2 \end{array}}
\q{\begin{array}{ccc} S & S & J \\ m_1' & m_2' & 
m_1'+ m_2' \end{array}}\\
&\quad\quad\times \delta_{m_1+m_2, m_1'+ m_2'}
|S; m_1' \rangle \langle S; m_1|
 \otimes
|S; m_2' \rangle \langle S;m_2| .
\end{split}
\end{align}

\begin{figure}
\begin{center}
\includegraphics[height=25mm,clip]{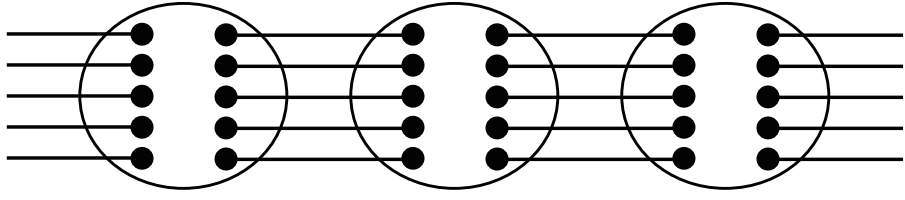}
\caption{Conceptual figure of the $q$-VBS state.
Each line is a $q$-deformed valence bond,
and the circle $\bigcirc$ represents the $q$-symmetrization of 
spin-1/2 particles $\bullet$ at each site.}
\label{q-VBS}
\end{center}
\end{figure}

The nonnegativity
\begin{align}\label{nonnegativity}
 \langle  \psi | \left(\pi_J\right)_{k,k+1} | \psi \rangle \ge 0
 \quad \left( \text{for any vector } | \psi \rangle \right)
\end{align}
implies that all the eigenvalues of $\mathcal H$
  are nonnegative.
(Of course, $\langle  \psi |$ is
 the Hermitian conjugate of $| \psi \rangle$.)
Moreover, we will see that the energy of
  the ground state $|\Psi\rangle$ is zero:
\begin{align}\label{HPsi=0}
  \mathcal H |\Psi\rangle =0.
\end{align}
Since we set $C_J(k,k+1)>0$,
 we find that \eqref{HPsi=0} is equivalent to
\begin{align}\label{local-condition}
   \left(\pi_J\right)_{k,k+1} | \Psi \rangle =0
    \quad \left(\forall k\in \Z_L,\ 
    \forall J\in \{ S+1,\dots,2S \} \right) ,
\end{align}
noting the nonnegativity \eqref{nonnegativity}.
From Proposition \ref{X^-^nv_J},
 one observes that any vector in
 $\bigoplus_{0\le J\le S}
 V_J \subset V_S \otimes V_S $
of the $k$-th and $(k+1)$-th sites has the form 
\begin{align}\label{local-form}
\sum_{0 \le A,B \le S} C_{A B}
x_k^{A} y_k^{S-A} x_{k+1}^{B} y_{k+1}^{S-B}
\prod_{m=1}^S
(q^m x_k y_{k+1} - q^{-m} y_k x_{k+1} )  ,
\end{align}
where $ C_{A B}$ does not depend on
 $x_k,y_k,x_{k+1}$ or $y_{k+1} $. 
Thus, the condition \eqref{local-condition}
  imposes the restriction that $|\Psi\rangle$
  has the form
\begin{align}
| \Psi \ket = P\left(\{x_k\}_{k\in\Z_L},
   \{y_k\}_{k\in\Z_L}\right)
\prod_{k\in\Z_L} \prod_{m=1}^S
(q^m x_k  y_{k+1} - q^{-m} y_k  x_{k+1} )
\end{align}
with some polynomials $P$
  such that this form is consistent with \eqref{local-form}
  for $\forall k\in\Z_L$.
The unique choice of $P$ with such consistency is a constant
   (which can be set to be 1),
  and we achieve the unique ground state
\begin{align}
| \Psi \ket =
\prod_{k\in\Z_L} \prod_{m=1}^S
(q^m x_k y_{k+1} - q^{-m} y_k  x_{k+1} ) .
\end{align}
In the Schwinger boson representation, we have
\begin{align}
| \Psi \ket=\prod_{k\in\Z_L} \prod_{m=1}^S
(q^m a_k^{\dagger} b_{k+1}^{\dagger}-q^{-m} b_k^{\dagger} a_{k+1}^{\dagger})
|\mathrm{vac} \ket,
\label{schwingergroundstate}
\end{align}
which is a generalization of the $q=1$ case \cite{AAH}.
Note that each site have the correct spin value:
$N_k| \Psi \rangle =S| \Psi \rangle$ ($k \in \Z_L$)
where $N_k:=(N_{a_k}+N_{b_k})/2$.
Our ground state is
  a $q$-deformation of the valence-bond-solid (VBS) state,
  which we call $q$-VBS state, see figure \ref{q-VBS}.

The Schwinger boson representation
of the ground state \eqref{schwingergroundstate} 
can be transformed into the following equivalent form
called the matrix product representation \cite{M}, which
generalizes the $q=1$ \cite{TS2} or $S=1$ \cite{KSZ1} case.
Noting \eqref{basis}, we have 
\begin{align}\label{mpf}
| \Psi \rangle&=\mathrm{Tr} 
  [g_1 \star g_2 \star \cdots \star g_{L-1} \star g_L],
\end{align}
where
$g_k$ is an $(S+1)\times(S+1)$
   vector-valued matrix 
acting on the $k$-th site
whose element is given by
\begin{align}
\begin{split}
g_k(i,i')
 &= (-1)^{S-i} q^{(i+ i'-S)(S+1)/2} 
    \sqrt{\left[\begin{array}{c} S \\ i \end{array} \right]
      \left[ \begin{array}{c} S \\ i' \end{array} \right] 
    [S-i+ i']! [S+i- i']!  }  \   |S; i'-i\rangle_k  \\
  &=: h_{i i'}  |S; i'-i  \rangle_k ,    \quad\quad (0\le i, i'\le S) .
\end{split}
\end{align}
The symbol $\star$ for two
   $(S+1)\times(S+1)$ vector-valued matrices
\begin{align}
  x=
   \left(\begin{array}{ccc}
      |x_{00}\rangle & \cdots &  |x_{0S}\rangle  \\
                \vdots & \ddots & \vdots  \\
      |x_{S0}\rangle & \cdots &  |x_{SS}\rangle  \\
   \end{array}\right)
  ,\quad
  y=
   \left(\begin{array}{ccc}
      |y_{00}\rangle & \cdots &  |y_{0S}\rangle  \\
                \vdots & \ddots & \vdots  \\
      |y_{S0}\rangle & \cdots &  |y_{SS}\rangle  \\
   \end{array}\right),
\end{align}
is defined by
\begin{align}
  x\star y=
   \left(\begin{array}{ccc}
      \sum_{u=0}^{S} |x_{0u}\rangle\otimes |y_{u0}\rangle   & \cdots  &  
      \sum_{u=0}^{S} |x_{0u}\rangle\otimes |y_{uS}\rangle   \\
                \vdots & \ddots & \vdots  \\
      \sum_{u=0}^{S} |x_{Su}\rangle\otimes |y_{u0}\rangle   & \cdots   &
      \sum_{u=0}^{S} |x_{Su}\rangle\otimes |y_{uS}\rangle   \\
   \end{array}\right),
\end{align}
which is associative.

For example, for $S=2$,
\begin{align}
  g_k  =
 \left(\begin{array}{ccc}
    h_{00} |2;0\rangle_k & h_{01} |2;1\rangle_k & h_{02} |2;2\rangle_k \\
    h_{10} |2;-1\rangle_k & h_{11} |2;0\rangle_k & h_{12} |2;1\rangle_k  \\
    h_{20} |2;-2\rangle_k & h_{21} |2;-1\rangle_k & h_{22} |2;0\rangle_k 
  \end{array}\right),
\end{align}
and the product in the form \eqref{mpf}
  is calculated as
\begin{align}
\begin{split}
  &  g_1 \star\cdots\star g_L \\
 =& \left(\begin{array}{ccc}
    \left( g_1 \star\cdots\star g_L \right) (0,0)  &
    \left( g_1 \star\cdots\star g_L\right) (0,1)  &
    \left( g_1 \star\cdots\star g_L\right) (0,2)  \\
    \left( g_1 \star\cdots\star g_L \right) (1,0)  &
    \left( g_1 \star\cdots\star g_L\right) (1,1)  &
    \left( g_1 \star\cdots\star g_L\right) (1,2)  \\
    \left( g_1 \star\cdots\star g_L \right) (2,0)  &
    \left( g_1 \star\cdots\star g_L\right) (2,1)  &
    \left( g_1 \star\cdots\star g_L\right) (2,2) 
  \end{array}\right),
\end{split}
\end{align}
with 
\begin{align}
\begin{split}
&  \left(  g_1 \star\cdots\star g_L  \right) (i,i')  \\
=& \sum_{i_k=0,1,2}h_{ii_2} h_{i_2i_3} \cdots h_{i_{L-1}i_L} h_{i_L i'} \\
&\quad \  \times
   |2;i_2-i\rangle_1 \otimes |2;i_3-i_2\rangle_2 \otimes\cdots\otimes 
   |2;i_L-i_{L-1}\rangle_{L-1} \otimes |2;i'-i_L\rangle_L .
\end{split}
\end{align}
Then the matrix product ground state \eqref{mpf} is
\begin{align}
\begin{split}
&\left( g_1 \star\cdots\star g_L \right) (0,0)+
\left( g_1 \star\cdots\star g_L \right) (1,1)+
\left( g_1 \star\cdots\star g_L \right) (2,2) \\
=
& \sum_{i_k=0,1,2}h_{i_1i_2} h_{i_2i_3} \cdots h_{i_{L-1}i_L} h_{i_Li_1} \\
&\ \times
   |2;i_2-i_1\rangle_1 \otimes |2;i_3-i_2\rangle_2 \otimes\cdots\otimes 
   |2;i_L-i_{L-1}\rangle_{L-1} \otimes |2;i_1-i_L\rangle_L .
\end{split}
\end{align}

We define $g_k^\dagger$
 by replacing each ket vector in the matrix $g_k$ 
 by its corresponding bra vector:
\begin{align}
   g^\dagger_k(i, i') = h_{i i'} \ _k \langle S; i'-i  |.
\end{align}
For example, for $S=2$, 
\begin{align}
  g^\dagger_k  =
 \left(\begin{array}{ccc}
    h_{00}\ _k \langle 2;0|  & h_{01}\ _k \langle 2;1|  &  h_{02}\ _k \langle 2;2|  \\
   h_{10}\ _k \langle 2;-1| &  h_{11}\ _k \langle 2;0| & h_{12}\ _k \langle 2;1|   \\
  h_{20}\ _k \langle 2;-2|   &  h_{21}\ _k \langle 2;-1|  &h_{22}\ _k \langle 2;0|    
\end{array}\right).
\end{align}
Now we introduce ``$G$ matrix'',
   which will play an important role in our study.
Let us set an $(S+1)^2$ dimensional vector space $W$ 
  and its dual orthogonal space $W^*$ as
\begin{align}
  W=\bigoplus_{0 \le a,b \le S} \C|a,b  \ketket  ,  \quad
  W^*=\bigoplus_{0 \le a,b \le S} \C\brabra a,b | .  \quad
\end{align}
Here,
$\{|a,b  \ketket \ | \ a,b=0, \dots, S  \}$
($\{ \brabra  a,b | \ | \ a,b=0, \dots, S  \}$)
is an orthonormal (dual orthonormal) basis.
We define an $(S+1)^2 \times (S+1)^2$
   matrix $G$ acting on the space $W$ as
\begin{align}
  G_{(a,b;c,d)} &=\brabra a,b | G | c,d \ketket =
  g^\dagger (a,c) g(b,d) ,
\label{gmatrix} 
\end{align}
or equivalently as
\begin{align}
  G=g^\dagger \otimes g .
\end{align}
We  also introduce $G_A$ for an operator $A$
 acting on the one-site vector space $V_S$  as
\begin{align}
   \left(G_A\right)_{(a,b;c,d)} &= 
   \brabra a,b | G_A | c,d \ketket =
   g^\dagger (a,c) Ag(b,d).
\end{align}
Each element of the matrix $G$ can be expressed explicitly as
\begin{align}
G_{(a,b;c,d)} = 
  \delta_{c-a,d-b}T_{abcd} ,
\end{align}
where 
\begin{align}
\begin{split}
&T_{abcd} = h_{ac} h_{bd} = 
  (-1)^{a+b} q^{(a+b+c+d-2S)(S+1)/2}  \\
&\ \times
\sqrt{
  \left[ \begin{array}{c} S \\ a \end{array} \right]
  \left[ \begin{array}{c} S \\ b \end{array} \right]
  \left[ \begin{array}{c} S \\ c \end{array} \right]
  \left[ \begin{array}{c} S \\ d \end{array} \right]
    [S-a+c]! [S+a-c]! [S-b+d]! [S+b-d]! }\ .
\end{split}
\end{align}
Each element of $G_A$ for $A=S^z, S^+$ and $S^-$,
 which act on $|S;m\rangle$ as
\begin{align}
  S^z |S;m\rangle  =& m |S;m\rangle,  \\
  S^+ |S;m\rangle =& \sqrt{(S-m)(S+m+1)} |S;m+1\rangle,  \\
  S^- |S;m\rangle  =& \sqrt{(S+m)(S-m+1)} |S;m-1\rangle, 
\end{align}
can be also expressed as
\begin{align}
\left(G_{S^z}\right)_{(a,b;c,d)}=&\delta_{c-a,d-b}(d-b) T_{abcd}, 
   \label{GSZ}\\
\left(G_{S^+}\right)_{(a,b;c,d)}=&\delta_{c-a,d-b+1}
     \sqrt{(S-d+b)(S+d-b+1)}  T_{abcd}, 
  \label{GSPLUS} \\
\left(G_{S^-}\right)_{(a,b;c,d)}=&\delta_{c-a,d-b-1} 
     \sqrt{ (S+d-b)(S-d+b+1) } T_{abcd}
  \label{GSMINUS}.
\end{align}
The squared norm of the ground state is calculated as
\begin{align}
\begin{split}
  \langle\Psi|\Psi\rangle
   =& \Tr  \left[ g_1^\dagger \star \cdots \star g_L^\dagger \right]
       \Tr  \left[ g_1 \star\cdots\star g_L \right]  \\
 =&  \Tr  \left[\left( g_1^\dagger \star \cdots \star g_L^\dagger \right)
  \otimes \left( g_1 \star \cdots \star g_L \right) \right]  \\
 =& \Tr \left[\left( g_1^\dagger \otimes g_1 \right) \star 
   \cdots \star \left( g_L^\dagger \otimes g_L \right) \right] \\
 =& \Tr G^L. 
\label{PsiPsi=TrGL}
\end{split}
\end{align}
Note that the elements of $g^\dagger_k \otimes g_k = G$
   are no longer vectors,
   and thus we can replace the symbol $\star$ by the usual product
   in the third line of \eqref{PsiPsi=TrGL}.
The one point function $ \langle A  \rangle$
 of an operator $A$ can be written 
 in terms of $G$ and $G_A$ as
\begin{align}
\begin{split}
 \langle A  \rangle
  =\frac{\langle\Psi|A_1|\Psi\rangle}{\langle\Psi|\Psi\rangle}
  =\frac{\Tr \left[g_1^\dagger \star \cdots \star g_L^\dagger\right]
       \Tr  \left[A_1g_1 \star g_2 \star  \cdots \star g_L           \right]  }
       {\langle\Psi|\Psi\rangle}
  = \frac{\Tr G_A G^{L-1} }{ \Tr G^L }, \label{onepoint} 
\end{split}
\end{align}
where $A_kg_k$ is defined by
 $\left(A_kg_k\right)(i, i')=A_k\left(g_k(i, i')\right)$.
In the same way, the two point function of $A$ and $B$
can also be written in terms of $G, G_A$ and $G_B$ as
\begin{align}
 \langle A_1 B_r  \rangle& = \left(\Tr G^L\right)^{-1} 
\Tr G_A G^{r-2} G_B G^{L-r}. \label{twopoint}
\end{align}

Investigating the eigenvalues and eigenvectors of the matrix $G$
   will be crucial for the analysis of correlation functions.
In the next section, we study the $G$ matrix in detail.

\section{\label{Sec:spec-of-G} Spectral structure of the $G$ matrix}

In Ref.~\cite{M}, we conjectured that the spectrum of $G$ is given by
\begin{align}
 \lambda_\ell=(-1)^\ell\left(\q{S}!\right)^2
      \q{\begin{array}{c} 2S+1 \\ S-\ell \end{array}},
  \quad(\ell=0,1,\dots,S), \label{eigenvalues}
\end{align}
where the degree of the degeneracy of each $\lambda_\ell$
is $2\ell+1$.
One can easily find that
\begin{align}\label{0>1>cdots>S}
   |\lambda_0|  >  |\lambda_1|>  \cdots >|\lambda_S| .
\end{align}
In this section, we prove the conjecture
by giving an exact form for the eigenvector
corresponding to each eigenvalue. \\
First one observes that the
$G$ matrix has the following block diagonal structure:
\begin{align}
 G=&\bigoplus_{-S\le j\le S}G^{ (j) },\quad G^{ (j) }\in {\rm End} W_j,\\
 W=& \bigoplus_{-S\le j\le S}W_j,  \quad 
  W_j=\begin{cases}
\displaystyle
\bigoplus_{0\le i \le S-j}\C|i,i+j\ketket & j\ge 0,\\
\displaystyle
\bigoplus_{0\le i \le S+j}\C|i-j,i \ketket & j<0.
\end{cases}
\end{align}
The size of each block $G^{ (j) }$ is $(S-|j|+1)\times(S-|j|+1)$.
Each element of $G^{ (j) }$ is 
\begin{align}
\begin{split}
 \brabra a,a+j|G^{ (j) }|c,c+j\ketket
 =&(-1)^jq^{(a+c+j-S)(S+1)}   \q{S-a+c}!\q{S+a-c}! \\
  & \times \sqrt{
     \q{\begin{array}{c} S \\ a \end{array}}
     \q{\begin{array}{c} S \\ a+j \end{array}}
     \q{\begin{array}{c} S \\ c \end{array}}
     \q{\begin{array}{c} S \\ c+j \end{array}}
      } \ .
\end{split}
\end{align}
We construct intertwiners among the $2S+1$ blocks
 $G^{ (j) }\ (j=-S, \dots, S) $.
This helps us to construct
eigenvectors of each block from
 another block with a smaller size.
(The same idea was used in Ref.~\cite{AKSS} to study
the spectrum of a multi-species exclusion process).
Let us define a family of linear operators
$\{I_j\}_{-S\le j\le-1,1\le j\le S}$ as
\begin{align}
I_j&\in 
{\rm Hom}(W_j,W_{j-1}), \\
\brabra a,a+j-1 |I_j| c,c+j \ketket
 &=
  \begin{cases}
  \displaystyle
  q^{-a}\sqrt{  \frac{\q{a+j}\q{S-a-j+1}}{\q{j}\q{S-j+1}}  } & c=a, \\
  \displaystyle
  -q^{1-a-j}\sqrt{ \frac{\q{a} \q{S-a+1}}{\q{j}\q{S-j+1}}  } & c=a-1, \\
  0                                   & \rm otherwise 
 \end{cases}  \\
\nonumber
\end{align}
for $1\le j\le S$, and
\begin{align}
I_j&\in {\rm Hom}(W_j,W_{j+1}), \\
\brabra a-j-1,a |I_j| c-j,c \ketket
 &=
  \begin{cases}
  \displaystyle
  q^{-a}\sqrt{ \frac{\q{a-j}\q{S-a+j+1}}{\q{-j}\q{S+j+1}} } & c=a, \\
  \displaystyle
  -q^{1-a+j}\sqrt{ \frac{\q{a} \q{S-a+1}}{\q{-j}\q{S+j+1}} } & c=a-1, \\
  0                                   & \rm otherwise .
 \end{cases}\\
\nonumber
\end{align}
for $-S\le j\le -1$.
By direct calculation, one finds
\begin{proposition}
The matrix $I_j$ enjoys the intertwining relation
\begin{align}
\begin{split}
\label{intrel}
 I_j G^{ (j) } =& G^{ (j-1) } I_j \quad {\rm for }\ 1\le j\le S,\\
 I_j G^{ (j) } =& G^{ (j+1) } I_j \quad {\rm for }\ -S\le j\le -1.
\end{split}
\end{align}
\label{intertwining}
\end{proposition}
With the use of Proposition \ref{intertwining}, one can show the following:
\begin{theorem}\label{eigen-structure}
Each block $G^{ (j) }$ has
 a simple (nondegenerated) spectrum
\begin{align}
  {\rm Spec}\ G^{ (j) }= \{ \lambda_\ell \}_{|j| \le  \ell\le S},
\end{align}
and the corresponding eigenvectors
are given by
\begin{align}
    \label{edgeeigen}
|\lambda_{|j|}   \ketket_j&=
\begin{cases}
\displaystyle
  \sum_{0\le i\le S-\ell}q^{(\ell+1)i}
   \sqrt{\frac{\q{S-\ell}! \q{i+\ell}! \q{S-i}!}{
      [S]! [\ell]! [S -i-\ell]! [i]! } }
    |i,i+\ell\ketket & j\ge 0, \\
\displaystyle
  \sum_{0\le i\le S-\ell}q^{(\ell+1)i}
   \sqrt{\frac{\q{S-\ell}! \q{i+\ell}! \q{S-i}!}{
        [S]! [\ell]! [S -i-\ell]! [i]! } }
    |i+\ell,i\ketket & j<0,
\end{cases}
\end{align}
for $\ell=|j|$, and
\begin{align}
\label{nakaeigen}
|\lambda_\ell\ketket_j&=
 \begin{cases}
  I_{j+1}|\lambda_\ell \ketket_{j+1}
   =I_{j+1}I_{j+2}\cdots I_{\ell} |\lambda_\ell\ketket_\ell & j\ge0, \\
  I_{j-1}|\lambda_\ell \ketket_{j-1}
  =I_{j-1}I_{j-2}\cdots I_{-\ell} |\lambda_\ell\ketket_{-\ell} & j<0. \\
 \end{cases}
\end{align}
for $|j|+1\le \ell \le S$.
\label{gmatrixstructure}
\end{theorem}

\begin{figure}[h]
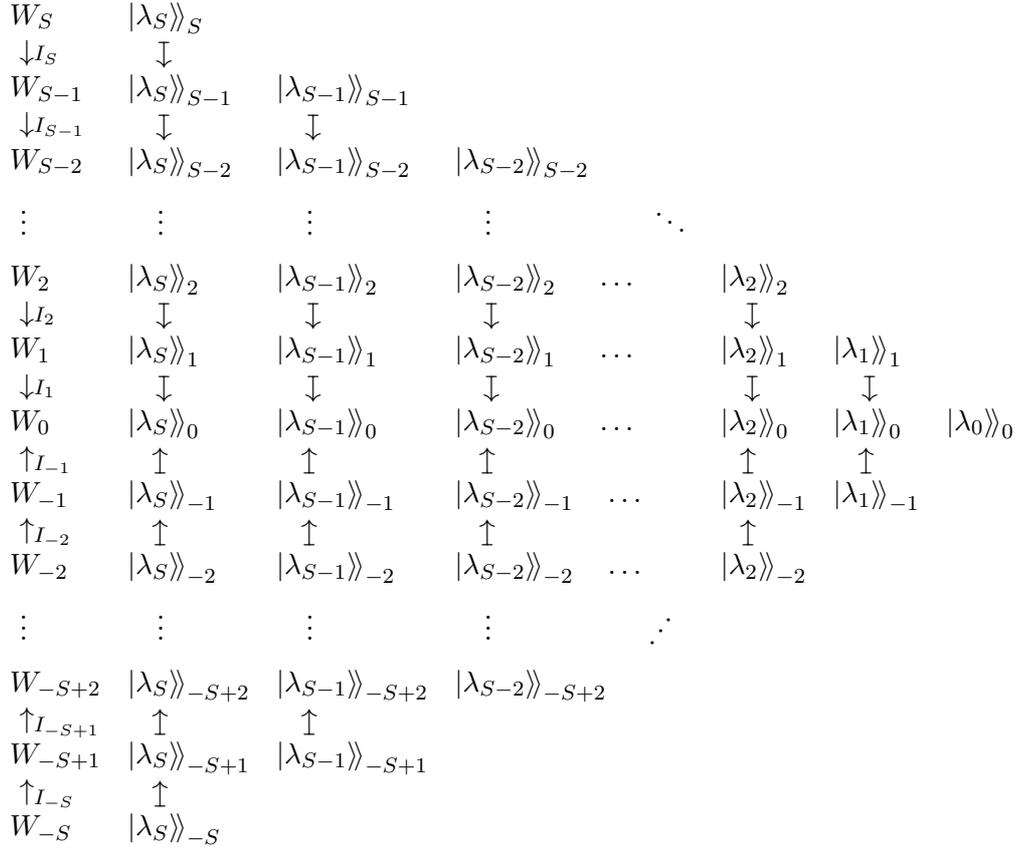

\begin{align*}
\begin{array}{llllllllll}
W_S    & |\lambda_S\ketket_S \\
\ \downarrow \scriptstyle{I}_S &  \quad\mapsdown  \\
W_{S-1}& |\lambda_S\ketket_{S-1} &
                      |\lambda_{S-1}\ketket_{S-1} \\
\ \downarrow\scriptstyle{I}_{S-1} &  \quad\mapsdown  & \quad\mapsdown \\
W_{S-2}& |\lambda_S\ketket_{S-2} &
         |\lambda_{S-1}\ketket_{S-2} &
    |\lambda_{S-2}\ketket_{S-2}
\vspace{2mm} \\
\ \vdots &  \quad\vdots & \quad \vdots
   & \quad\vdots \quad\quad\quad\quad\quad\ \ddots
\vspace{2mm} \\
W_2    & |\lambda_S\ketket_2  &|\lambda_{S-1}\ketket_2 &
    |\lambda_{S-2}\ketket_2 \quad \ \dots
   &\,|\lambda_2\ketket_2 \\
\ \downarrow\scriptstyle{I}_2 & \quad\mapsdown & \quad\mapsdown
      & \quad\mapsdown & \quad\mapsdown \\
W_1    & |\lambda_S\ketket_1  & |\lambda_{S-1}\ketket_1 &
    |\lambda_{S-2}\ketket_1 \quad \  \dots
   &\,|\lambda_2\ketket_1&|\lambda_1\ketket_1  \\
    \ \downarrow\scriptstyle{I}_1 & \quad\mapsdown & \quad\mapsdown
      & \quad\mapsdown & \quad\mapsdown & \quad\mapsdown& \\
W_0    &  |\lambda_S\ketket_0  &|\lambda_{S-1}\ketket_0 &
  |\lambda_{S-2}\ketket_0  \quad \  \dots
   &\, |\lambda_2\ketket_0 & |\lambda_1\ketket_0 &|\lambda_0\ketket_0   \\
\ \uparrow_{I_{-1}} & \quad\mapsup & \quad\mapsup
      & \quad\mapsup & \quad\mapsup & \quad\mapsup& \\
W_{-1}  & |\lambda_S\ketket_{-1}  & |\lambda_{S-1}\ketket_{-1} &
      |\lambda_{S-2}\ketket_{-1} \quad \dots
   &\, |\lambda_2\ketket_{-1}&|\lambda_1\ketket_{-1}   \\
\ \uparrow_{I_{-2}} & \quad\mapsup & \quad\mapsup
      & \quad\mapsup & \quad\mapsup \\
W_{-2}    & |\lambda_S\ketket_{-2}  &|\lambda_{S-1}\ketket_{-2} &
    |\lambda_{S-2}\ketket_{-2} \quad \dots
   &\, |\lambda_2\ketket_{-2}
\vspace{2mm} \\
\ \vdots &  \quad\vdots & \quad \vdots
   & \quad\vdots  \quad\quad\quad\quad\quad\  \dotsd
\vspace{2mm} \\
W_{-S+2}& |\lambda_S\ketket_{-S+2} &
         |\lambda_{S-1}\ketket_{-S+2} &
    |\lambda_{S-2}\ketket_{-S+2}  \\
\ \uparrow_{I_{-S+1}} &  \quad\mapsup  & \quad\mapsup \\
W_{-S+1}& |\lambda_S\ketket_{-S+1} &
                      |\lambda_{S-1}\ketket_{-S+1}  \\
\ \uparrow_{I_{-S}} &  \quad\mapsup \\
W_{-S} & |\lambda_S\ketket_{-S}  \\
\end{array}
\end{align*}
\caption{Structure of the eigenvectors of $G$ \eqref{gmatrix}.
}\label{structure}
\end{figure}

Figure \ref{structure} is helpful to understand
  how the eigenvectors are constructed.
We prove this theorem below
  for only $j\ge 0$
  since one can show it for $j<0$ in the same way.

{\it Proof of Theorem \ref{eigen-structure}.}
First, by direct calculation given below,
we find that
 $G^{(j)}$ has an eigenvalue $\lambda_j$
 and its eigenvector is
 $|\lambda_j\ketket_j$ defined by \eqref{edgeeigen}.
Each element of $ G^{ (j) }|\lambda_j \ketket_j $
  is calculated as
\begin{align}
\begin{split}
 &  \brabra a, a+j|G^{ (j) }|\lambda_j \ketket_j  \\
  &= \sum_{0\le c\le S-j}(-1)^j
       q^{(a+c+j-S)(S+1)}   \q{S-a+c}!\q{S+a-c}! \\
   & \quad\times \sqrt{
     \q{\begin{array}{c} S \\ a \end{array}}
     \q{\begin{array}{c} S \\ a+j \end{array}}
     \q{\begin{array}{c} S \\ c \end{array}}
     \q{\begin{array}{c} S \\ c+j \end{array}}   } 
   q^{(j+1)c}\displaystyle 
   \sqrt{\frac{\q{S-j}! \q{c+j}! \q{S-c}!}{
      \q{S}! \q{j}! \q{S-c-j}! \q{c}! } }
\\
  &= (-1)^j  q^{(a+j-S-1)(S+1)-(j+1)}
     \sqrt{
     \q{\begin{array}{c} S \\ a \end{array}}
     \q{\begin{array}{c} S \\ a+j \end{array}}
     \displaystyle \frac{\q{S-j}!}{\q{S}! \q{j}!}} \\
   &\quad\times  \q{S}!
    \sum_{0\le c\le S-j } q^{(c+1)(S+j+2)}
     \frac{\q{S-a+c}!\q{S+a-c}!}{\q{S-c-j}! [c]!} .
\end{split}
\end{align}
Using the formula
\begin{align}
\sum_{0\le k\le n}
\q{
\begin{array}{c} \alpha+n-k \\ n-k \end{array} } 
\q{ \begin{array}{c} \beta+k \\ k \end{array} } 
q^{k(\alpha+\beta+2)}
=\q{ \begin{array}{c} \alpha+\beta+n+1 \\ n \end{array}  }
q^{n(1+\beta)},
\end{align}
we obtain
\begin{align}
\begin{split}
\brabra a, a+j|G^{ (j) }|\lambda_j \ketket_j 
  &= (-1)^j q^{(a+j-S-1)(S+1)-(j+1)}
     \sqrt{
     \q{\begin{array}{c} S \\ a \end{array}}
     \q{\begin{array}{c} S \\ a+j \end{array}}
     \displaystyle \frac{\q{S-j}!}{\q{S}! \q{j}!}} \\
   &\quad\times  \q{S}!\,
     q^{-Sj+a(j-S)+S^2+2S+2}
     \q{\begin{array}{c} 2S+1 \\ S-j \end{array}}
     \q{S-a}!\q{a+j}!
\\
  &= (-1)^j\left(\q{S}!\right)^2
       \q{\begin{array}{c} 2S+1 \\ S-j \end{array}}
       q^{(j+1)a}
     \sqrt{\displaystyle
     \frac{\q{S-j}!\q{a+j}!\q{S-a}!}{\q{S}!\q{j}!\q{S-a-j}!\q{a}!} }
\\
  &=  \lambda_j \brabra a, a+j |\lambda_j\ketket_j .
\end{split}
\end{align}
Note that the first element of $|\lambda_j\ketket_j$ is 1
  by the definition \eqref{edgeeigen}:
   $ \brabra 0,j | \lambda_j\ketket_j=1$.

Next, we show by induction that
  $G^{(j)}$ has eigenvalues $\lambda_\ell$ ($j\le\ell\le S$)
  and their corresponding eigenvectors are
  given by  $|\lambda_\ell\ketket_j$ defined by \eqref{nakaeigen}. 
Suppose the theorem is true for $|\lambda_\ell  \ketket_{j+1},
\ell=j+1, \dots,S \ (j \ge 0)$,
 that is to say that the block
 $G^{ (j+1) }$ has the eigenvalues $\lambda_\ell$ and their
corresponding eigenvectors $| \lambda_\ell  \ketket_{j+1}$
$\big( G^{ (j+1) }|\lambda_\ell\ketket_{j+1}=
\lambda_\ell|\lambda_\ell\ketket_{j+1} $
with $|\lambda_\ell\ketket_{j+1} \neq 0\big)$
 for $\ell=j+1,\dots,S$.
Additionally, suppose
 that the first element of each $|\lambda_\ell\ketket_{j+1} $ is 1.
Using the intertwining relation \eqref{intrel},
one finds $G^{ (j) }I_{j+1}|\lambda_\ell\ketket_{j+1}
   =\lambda_\ell I_{j+1}|\lambda_\ell\ketket_{j+1}$.
We also find that the first element of
  $ I_{j+1}|\lambda_\ell\ketket_{j+1} $ is 1,
  and thus $I_{j+1}|\lambda_\ell\ketket_{j+1}$ is nonzero.
Furthermore, thanks to
$\ell_1\neq\ell_2\Rightarrow\lambda_{\ell_1} \neq\lambda_{\ell_2}$,
the vectors $I_{j+1}|\lambda_\ell\ketket_{j+1}$ (${j+1\le\ell\le S}$)
are distinct (in other words, $I_{j+1}$ is injective).
We have already constructed the remaining 
eigenvector of $G^{(j)}$ explicitly, which is
$| \lambda_j  \ketket_j$ with
its eigenvalue $\lambda_j$ distinct from 
$\lambda_\ell\ (j+1\le\ell\le S)$.
\qed

The conjecture for the eigenvalues
of the $G$ matrix that we exhibited in the beginning of this section
follows as a simple corollary of Theorem \ref{gmatrixstructure}.
Moreover, we constructed their eigenvectors
which are important for computing spin-spin correlation functions.

\begin{proposition}\label{prop-squared-norm}
  The squared norm of $|\lambda_\ell\ketket_j$ is
 \begin{align}
  {}_j\brabra \lambda_\ell|\lambda_\ell\ketket_j
  =q^{S(|j|+1) - \ell(\ell+1) }
   \frac{  [S+\ell+1]!  [\ell-|j|]!  [S-\ell]!  [|j|]!  }
  { [S]!  [\ell+|j|]!  [S-|j|]!  [2\ell+1] },
    \label{squared-norm}
 \end{align}
where we denote the transpose of $| \lambda_\ell  \ketket_j$ by
 ${}_j  \brabra \lambda_\ell|$.
\end{proposition}
We prove this proposition only for $j\ge 0$.

{\it Proof of Proposition \ref{prop-squared-norm}. }
One can easily show
  that the product of intertwiners 
  (which is also an intertwiner)
   has the following form by induction:
\begin{align}
\begin{split}
&\brabra  a,a+j |
 I_{j+1} I_{j+2}\cdots I_{\ell+1}I_{\ell}
  | c,c+\ell \ketket \\
=&
 (-1)^{a - c} q^{c j - a \ell}
 \q{\begin{array}{c} \ell-j \\  a - c\end{array}  } 
\sqrt{ 
  \frac{  [j]! [S-\ell]! [a]!  [S-c]! [c+\ell]! [S-(a+j) ]!  }
      { [\ell]! [S-j]! [c]!  [S-a]! [a+j]!  [ S -(c + \ell) ]!  }   }  .
\end{split}
\end{align}
Then,
 ${}_j  \brabra \lambda_\ell |\lambda_\ell\ketket_j
    ={}_\ell  \brabra \lambda_\ell  | 
   \left(I_{j+1} \cdots I_\ell\right)^{\rm T}
   I_{j+1} \cdots I_\ell |\lambda_\ell\ketket_\ell $
is calculated as
\begin{align}
\begin{split}
&{}_j  \brabra \lambda_\ell |\lambda_\ell\ketket_j
=   \frac{\left([S-\ell]!\right)^2 [j]! }{ [S]! \left([\ell]!\right)^2 [S-j]! }
  \sum_{0 \le a\le S-j \atop 0\le i,i'\le S }
    (-1)^{i+i'} q^{( i+i')(\ell+j+1)-2a\ell}  \\
& \quad\quad \times 
  \q{ \begin{array}{c} \ell-j \\ a-i \end{array} }
  \q{ \begin{array}{c} \ell-j \\ a-i' \end{array} }
\frac{[S-i]! [i+\ell]! [S-i']! [i'+\ell]! [a]! [S-(a+j)]! }
{[i]! [S-(i+\ell)]! [i']! [S-(i'+\ell)]! [S-a]![a+j]!} .
\end{split}
\end{align}
The triple sum has the closed form
\begin{align}
  q^{(j+1)S-\ell(\ell+1)}
 \frac{ \left([\ell]!\right)^2  [\ell-j]![S+\ell+1]!}
  {[S-\ell]! [j+\ell]! [2\ell+1]},
\end{align}
which finishes the proof.\qed

\section{\label{Sec:correlation}Spin-spin correlation functions}
In the last section, we investigated the
eigenvalues and eigenvectors of the $G$ matrix.
By utilizing Theorem \ref{gmatrixstructure}
 and noting \eqref{0>1>cdots>S},
the one point function $\langle A  \rangle$
 can be represented as
\begin{align}
 \langle A  \rangle=\lambda_0^{-1} 
\frac{ {}_0  \brabra \lambda_0 | G_A | \lambda_0   \ketket_0}
{{}_0  \brabra \lambda_0 |\lambda_0   \ketket_0}
\end{align}
in the thermodynamic limit $L\to\infty$.
As an application, we can calculate
the probability of finding $S^z=m$ value as
\begin{align}
\begin{split}
 {\rm Prob}(S^z=m)
 =& \big\langle\,   |S;m \rangle \langle S;m | \,  \big\rangle  \\
 =& \frac{[S+m]![S-m]!}{[2S+1]!}
\sum_{i=0}^{S} q^{(S+2)(2i-m-S)}
\left[
\begin{array}{c}
S \\
i-m
\end{array}
\right]
\left[
\begin{array}{c}
S \\
i
\end{array}
\right].
\end{split}
\end{align}
The two point function \eqref{twopoint} can be also represented as
\begin{align}
\langle A_1 B_r  \rangle=\sum_{\ell=0}^S \lambda_\ell^{-2}
\left( \frac{\lambda_\ell}{\lambda_0} \right)^r
\sum_{j=-\ell}^\ell
 \frac{{}_0  \brabra  \lambda_0 |G_A | \lambda_\ell  \ketket_j
{}_j  \brabra  \lambda_\ell |G_B | \lambda_0  \ketket_0}
{{}_0  \brabra  \lambda_0 | \lambda_0  \ketket_0
{}_j  \brabra  \lambda_\ell | \lambda_\ell  \ketket_j},
 \label{twopointrep}
\end{align}
in the thermodynamic limit.
Inserting
\eqref{GSZ}, \eqref{GSPLUS}, \eqref{GSMINUS}, \eqref{eigenvalues},
    \eqref{edgeeigen}, \eqref{nakaeigen} and \eqref{squared-norm}
into \eqref{twopointrep}, one finds
 the large-distance ($r\to \infty$) behaviors
 of the spin-spin correlation functions
 $ \langle S_1^z S_r^z  \rangle$
 and $ \langle S_1^+ S_r^-  \rangle $ are
\begin{align}
 \langle S_1^z S_r^z  \rangle&=-
\frac{[3]  [S+2] }{q^{2S-2}
[S]  ([2S+1] !)^2}
({}_0  \brabra \lambda_1|G_{S^z}| \lambda_0  \ketket_0)^2
\left(-\frac{[S] }{[S+2] } \right)^r, \\
 \langle S_1^+ S_r^-  \rangle&=-
\frac{[2]  [3]  [S+2] }{q^{3S-2} ([2S+1] ! [S] )^2}
({}_{-1}  \brabra \lambda_1|G_{S^-}| \lambda_0  \ketket_0)^2
\left(-\frac{[S] }{[S+2] } \right)^r,
\end{align}
where
\begin{align}
\begin{split}
& \!\!\!\!\!\!
{}_0  \brabra \lambda_1|G_{S^z}| \lambda_0  \ketket_0
=\frac{q^{-S^2-S-1}}{q^S-q^{-S} }
  \sum_{i, i'=0}^{S}(i- i')q^{(S+2)(i+ i')}  \\
  &\times \{q^{S+1}+q^{-S-1}-(q+q^{-1})q^{2 i'-S} \} 
  [S+i- i'] ! [S+ i'-i] ! 
\left[ \begin{array}{c} S \\ i \end{array} \right] 
\left[ \begin{array}{c} S \\ i' \end{array} \right] ,
\end{split}
\\
\begin{split}
& \!\!\!\!\!\!
{}_{-1}  \brabra \lambda_1|G_{S^-}| \lambda_0  \ketket_0 =
{}_0     \brabra \lambda_0|G_{S^+}| \lambda_1  \ketket_{-1} \\
&=-q^{-S^2-S/2+1/2} \sum_{i=0}^{S} \sum_{ i'=0}^{S-1}
q^{(S+2)i+(S+3) i'}
 \sqrt{\left[\begin{array}{c} S \\ i'+1 \end{array} \right]
      \left[ \begin{array}{c} S \\ i' \end{array} \right]} 
\\
&\quad \times 
\sqrt{(S+i- i') [S+i- i']  (S-i+ i'+1) [S-i+ i'+1]}
\\
&\quad \times
\sqrt{ [i'+1][S-i'] [S]^{-1} } [S+ i'-i] ! [S+i- i'-1] !
\left[ \begin{array}{c} S \\i \end{array}\right] .
\end{split}
\end{align}
Note that the terms with $(j,\ell)=(0,1)$ and $(-1,1)$
 in \eqref{twopointrep} dominate
 the large-distance behaviors of $\langle S^z_1 S^z_r \rangle$
 and $\langle S^+_1 S^-_r \rangle$, respectively,
since
\begin{align}
  {}_0\brabra\lambda_0|G_{S^z}|\lambda_0\ketket_0
 ={}_1\brabra \lambda_1|G_{S^z}|\lambda_0\ketket_0
 ={}_{-1}\brabra \lambda_1|G_{S^z}|\lambda_0\ketket_0 =0, \\
  {}_0\brabra\lambda_0|G_{S^-}|\lambda_0\ketket_0
 ={}_1\brabra \lambda_1|G_{S^-}|\lambda_0\ketket_0
 ={}_0\brabra \lambda_1|G_{S^-}|\lambda_0\ketket_0 =0.
\end{align}
Both $ \langle S_1^z S_r^z  \rangle$ and 
$ \langle S_1^+ S_r^-  \rangle$ exhibit exponential decay with
correlation length
\begin{align}
\zeta=\left( \ln \frac{[S+2] }{[S] } \right)^{-1},
\end{align}
generalizing the results for $q=1$ \cite{FH}
or $S=1$ \cite{KSZ1} case.

\section{Conclusion \label{Sec:Conclusion}}
In this paper, we investigated one and two point functions
of the $q$-VBS ground state of
an integer spin model (the $q$-deformed higher-spin AKLT model).
The formulation of correlation functions by use of the matrix
product representation of the ground state shows that the structure
of a matrix, which we call $G$ matrix, plays an important role.
We obtained the eigenvalues and eigenvectors of the 
$G$ matrix with the help of constructing intertwiners connecting
different blocks of $G$.
Then we calculated the spin-spin correlation functions
by use of the eigenvalues and eigenvectors of the $G$ matrix,
and determined the correlation amplitudes and correlation lengths of the
longitudinal and transverse spin-spin correlation functions.

It is interesting to investigate other types of correlation functions.
For example, the entanglement entropy,
 which is defined in terms of the reduced density matrix,
 is a typical quantification of the entanglement
 of quantum systems.
It is intriguing to calculate the entanglement entropy
for the $q$-deformed model and observe the change from the isotropic point
\cite{XKHK,FKR,KHH} (see also Refs.~\cite{KHK,KKKKT}
 for other VBS states).

\section*{Acknowledgements}
The authors thank Atsuo Kuniba and Kazumitsu Sakai
 for useful discussion.
CA also thanks Kirone Mallick and Andreas Schadschneider
  for the very kind hospitality during his stay in Europe. 
This work is supported by Grant-in-Aid for Young Scientists (B) 22740106
  and Global COE program
 ``Education and Research Hub for Math-for-Industry.''

\appendix

\section{\label{Sec:proof}Proof of Proposition \ref{X^-^nv_J} }
We prove Proposition \ref{X^-^nv_J} by induction.
Suppose that Proposition
\ref{X^-^nv_J} holds for $\left( \Delta X^-_{\alpha\beta} \right)^n v_J$.
We calculate the four terms of
  the action of
\begin{align}
\begin{split}
  \Delta X^-_{\alpha\beta}
  =& \frac{1}{q-q^{-1}}\frac{y_\alpha}{x_\alpha} D_q^{x_\alpha} \otimes
    D^{x_\beta}_{\sqrt{q}} D^{y_\beta}_{1/\sqrt{q}} 
  -\frac{1}{q-q^{-1}}\frac{y_\alpha}{x_\alpha} D^{x_\alpha}_{q^{-1}}  \otimes
    D^{x_\beta}_{\sqrt{q}} D^{y_\beta}_{1/\sqrt{q}}
  \\
  &+\frac{1}{q-q^{-1}}D^{x_\alpha}_{1/\sqrt{q}} D^{y_\alpha}_{\sqrt{q}}
     \otimes  \frac{y_\beta}{x_\beta}  D^{x_\beta}_q   
  -\frac{1}{q-q^{-1}}D^{x_\alpha}_{1/\sqrt{q}} D^{y_\alpha}_{\sqrt{q}}
     \otimes  \frac{y_\beta}{x_\beta}  D^{x_\beta}_{q^{-1}},
\end{split}
\end{align}
on $\left( \Delta X^-_{\alpha\beta} \right)^n v_J$, separately.

\begin{align}
\label{first}
&  \left(\frac{y_\alpha}{x_\alpha}  D_q^{x_\alpha}   \otimes
    D^{x_\beta}_{\sqrt{q}} D^{y_\beta}_{1/\sqrt{q}} \right)
    \left( \Delta X_{\alpha\beta}^- \right)^n v_J
     \bigg/ q^{(n+1)S}[n]! (x_\alpha x_\beta)^{J-(n+1)} 
\\ =& \nn
     \sum_{\mu=0}^n  q^{-2\mu S + J - n}
       \q{\begin{array}{c} J \\ \mu \end{array} }
       \q{\begin{array}{c} J \\ n-\mu \end{array} }
     \left(x_\alpha y_\beta \right)^\mu
     \left(x_\beta y_\alpha \right)^{n+1-\mu} 
     \prod_{\nu=1}^{2S-J}
     \left(x_\alpha y_\beta -q^{2(\nu-S-1)}x_\beta y_\alpha \right) ,
\end{align}

\begin{align}
\label{second}
&  \left(\frac{y_\alpha}{x_\alpha}  D^{x_\alpha}_{q^{-1}}  \otimes
    D^{x_\beta}_{\sqrt{q}} D^{y_\beta}_{1/\sqrt{q}} \right)
    \left( \Delta X_{\alpha\beta}^- \right)^n v_J
     \bigg/ q^{(n+1)S}[n]!(x_\alpha x_\beta)^{J-(n+1)} 
\\ =&  \nn
     \sum_{\mu=0}^n  q^{-2\mu S -2\mu - 4S  + J + n}
       \q{\begin{array}{c} J \\ \mu \end{array} }
       \q{\begin{array}{c} J \\ n-\mu \end{array} }
     \left(x_\alpha y_\beta \right)^\mu
     \left(x_\beta y_\alpha \right)^{n+1-\mu} 
     \prod_{\nu=1}^{2S-J}
     \left(x_\alpha y_\beta -q^{2(\nu-S)}x_\beta y_\alpha \right),
\end{align}

\begin{align}
\label{third}
&  \left(
   D^{x_\alpha}_{1/\sqrt{q}} D^{y_\alpha}_{\sqrt{q}}
     \otimes  \frac{y_\beta}{x_\beta}  D_q^{x_\beta}     \right)
    \left( \Delta X_{\alpha\beta}^- \right)^n v_J
     \bigg/ q^{(n+1)S}[n]!(x_\alpha x_\beta)^{J-(n+1)} 
\\ =&    \nn
     \sum_{\mu=0}^n  q^{-2\mu S - 2\mu -2S + J + n }
       \q{\begin{array}{c} J \\ \mu \end{array} }
       \q{\begin{array}{c} J \\ n-\mu \end{array} }
     \left(x_\alpha y_\beta \right)^{\mu+1}
     \left(x_\beta y_\alpha \right)^{n-\mu} 
     \prod_{\nu=1}^{2S-J}
     \left(x_\alpha y_\beta -q^{2(\nu-S)}x_\beta y_\alpha \right),
\end{align}

\begin{align}
\label{fourth}
&  \left(
   D^{x_\alpha}_{1/\sqrt{q}} D^{y_\alpha}_{\sqrt{q}}
     \otimes  \frac{y_\beta}{x_\beta}  D^{x_\beta}_{q^{-1}}  \right)
    \left( \Delta X_{\alpha\beta}^- \right)^n v_J
   \bigg/ q^{(n+1)S}[n]!(x_\alpha x_\beta)^{J-(n+1)} 
\\ =&    \nn
     \sum_{\mu=0}^n  q^{-2\mu S -2S  - J + n}
       \q{\begin{array}{c} J \\ \mu \end{array} }
       \q{\begin{array}{c} J \\ n-\mu \end{array} }
     \left(x_\alpha y_\beta \right)^{\mu+1}
     \left(x_\beta y_\alpha \right)^{n-\mu} 
     \prod_{\nu=1}^{2S-J}
     \left(x_\alpha y_\beta -q^{2(\nu-S-1)}x_\beta y_\alpha \right).
\end{align}

\eqref{third} $-$ \eqref{second} gives
\begin{align}
&
\begin{split}
&
     \sum_{\mu=0}^n  q^{-2\mu S - 2\mu -2S + J + n }
       \q{\begin{array}{c} J \\ \mu \end{array} }
       \q{\begin{array}{c} J \\ n-\mu \end{array} }
     \left(x_\alpha y_\beta \right)^{\mu}
     \left(x_\beta y_\alpha \right)^{n-\mu} 
\\ & \quad \times
    \left( x_\alpha y_\beta -q^{-2S} x_\beta y_\alpha \right)
     \prod_{\nu=1}^{2S-J}
     \left(x_\alpha y_\beta -q^{2(\nu-S)} x_\beta y_\alpha \right)
\end{split}
\nonumber
\\ &
\begin{split}
=&     \sum_{\mu=0}^n  q^{-2\mu S - 2\mu -2S + J + n }
       \q{\begin{array}{c} J \\ \mu \end{array} }
       \q{\begin{array}{c} J \\ n-\mu \end{array} }
     \left(x_\alpha y_\beta \right)^{\mu}
     \left(x_\beta y_\alpha \right)^{n-\mu} 
\\ & \quad \times
    \left( x_\alpha y_\beta -q^{2(S-J)} x_\beta y_\alpha \right)
     \prod_{\nu=1}^{2S-J}
     \left(x_\alpha y_\beta -q^{2(\nu-S-1)} x_\beta y_\alpha \right)
\end{split}
\nonumber
\\ &
\begin{split}
=&     \sum_{\mu=0}^n  q^{-2\mu S - 2\mu -2S + J + n }
       \q{\begin{array}{c} J \\ \mu \end{array} }
       \q{\begin{array}{c} J \\ n-\mu \end{array} }
     \left(x_\alpha y_\beta \right)^{\mu+1}
     \left(x_\beta y_\alpha \right)^{n-\mu} 
\\ & \quad \times
     \prod_{\nu=1}^{2S-J}
     \left(x_\alpha y_\beta -q^{2(\nu-S-1)} x_\beta y_\alpha \right)
\\ & -
     \sum_{\mu=0}^n  q^{-2\mu S - 2\mu - J + n }
       \q{\begin{array}{c} J \\ \mu \end{array} }
       \q{\begin{array}{c} J \\ n-\mu \end{array} }
     \left(x_\alpha y_\beta \right)^{\mu}
     \left(x_\beta y_\alpha \right)^{n+1-\mu} 
\\ & \quad \times
     \prod_{\nu=1}^{2S-J}
     \left(x_\alpha y_\beta -q^{2(\nu-S-1)} x_\beta y_\alpha \right).
\label{third-second}
\end{split}
\end{align}
Dividing the first term of \eqref{third-second}  $-$ \eqref{fourth}
   by $q-q^{-1}$,  we obtain
\begin{align}
\begin{split}
&  \sum_{\mu=0}^n  q^{-2\mu S - \mu -2S  + n } [J-\mu]
       \q{\begin{array}{c} J \\ \mu \end{array} }
       \q{\begin{array}{c} J \\ n-\mu \end{array} }
     \left(x_\alpha y_\beta \right)^{\mu+1}
     \left(x_\beta y_\alpha \right)^{n-\mu} 
\\ & \quad \times
     \prod_{\nu=1}^{2S-J}
     \left(x_\alpha y_\beta -q^{2(\nu-S-1)} x_\beta y_\alpha \right)
\end{split}
\nonumber
\\
\begin{split}
=&  \sum_{\mu=1}^{n+1}  q^{-2\mu S  - \mu + n +1 } [J-\mu+1]
       \q{\begin{array}{c} J \\ \mu-1 \end{array} }
       \q{\begin{array}{c} J \\ n+1-\mu \end{array} }
     \left(x_\alpha y_\beta \right)^{\mu}
     \left(x_\beta y_\alpha \right)^{n+1-\mu} 
\\ & \quad \times
     \prod_{\nu=1}^{2S-J}
     \left(x_\alpha y_\beta -q^{2(\nu-S-1)} x_\beta y_\alpha \right)  ,
\label{mochotto1}
\end{split}
\end{align}
where we replaced $\mu\to\mu-1$.
Dividing \eqref{first}  $-$ the second term of \eqref{third-second}
 by $q-q^{-1}$, we obtain
\begin{align}
\begin{split}
&  \sum_{\mu=0}^n  q^{-2\mu S - \mu  } [J+\mu-n]
       \q{\begin{array}{c} J \\ \mu \end{array} }
       \q{\begin{array}{c} J \\ n-\mu \end{array} }
     \left(x_\alpha y_\beta \right)^{\mu}
     \left(x_\beta y_\alpha \right)^{n+1-\mu} 
\\ & \quad \times
     \prod_{\nu=1}^{2S-J}
     \left(x_\alpha y_\beta -q^{2(\nu-S-1)} x_\beta y_\alpha \right) .
\label{mochotto2}
\end{split}
\end{align}
Since $\q{\begin{array}{c} J \\ -1\end{array}}=0$,
  we can extend $\sum$ in \eqref{mochotto1} and \eqref{mochotto2}
  to $0\le \mu \le n+1$.
Finally we have
\begin{align}
\begin{split}
  \eqref{mochotto1} +  \eqref{mochotto2}
=&  \sum_{\mu=0}^{n+1}  q^{-2\mu S  } [n+1]
       \q{\begin{array}{c} J \\ \mu \end{array} }
       \q{\begin{array}{c} J \\ n+1-\mu \end{array} }
     \left(x_\alpha y_\beta \right)^{\mu}
     \left(x_\beta y_\alpha \right)^{n+1-\mu} 
\\ & \quad \times
     \prod_{\nu=1}^{2S-J}
     \left(x_\alpha y_\beta -q^{2(\nu-S-1)} x_\beta y_\alpha \right),
\end{split}
\label{final}
\end{align}
and since
\begin{align}
\begin{split}
 \left( \Delta X_{\alpha\beta}^- \right)^{n+1} v_J
     \bigg/ q^{(n+1)S}[n]! (x_\alpha x_\beta)^{J-(n+1)}
&=\frac{\eqref{first}+\eqref{third}-\eqref{second}-\eqref{fourth}}{q-q^{-1}}
\\
&=\frac{\eqref{first}+\eqref{third-second}-\eqref{fourth}}{q-q^{-1}}
\\
&=\eqref{mochotto1} +  \eqref{mochotto2} \\
&=\eqref{final}, 
\end{split}
\end{align}
Proposition \ref{X^-^nv_J} is true for 
$\left( \Delta X^-_{\alpha\beta} \right)^{n+1} v_J$.

\end{document}